\documentclass[aps,pra,onecolumn,8pt,showpacs]{revtex4}
\usepackage{mathrsfs}
\usepackage{graphicx}
\usepackage{amsmath}
\usepackage{amssymb}
\usepackage{CJK}
\usepackage{hyperref}
\usepackage{subfigure}

\begin{document}
\begin{CJK*}{GBK}{song}
\title{Scalable quantum information transfer between
nitrogen-vacancy-center ensembles}

\author{Feng-yang Zhang$^{1,3}$\footnote{zhangfy1986@gmail.com}, Chui-Ping Yang$^2$, and He-Shan Song$^3$}
\affiliation{$^1$School of Physics and Materials Engineering, Dalian Nationalities University, Dalian $116600$, China\\
$^2$Department of Physics, Hangzhou Normal University, Hangzhou, Zhejiang 310036, China\\
$^3$School of Physics
and Optoelectronic Technology, Dalian University of Technology, Dalian 116024, China}


\begin{abstract}
 We propose an architecture for realizing quantum information transfer (QIT). In this architecture, a \emph{LC} circuit is used to induce the necessary interaction between flux qubits, each magnetically coupling to a nitrogen-vacancy center ensemble (NVCE). We explicitly show that for resonant interaction and large detuning cases, high-fidelity QIT between two spatially-separated NVCEs can be implemented. Our proposal can be extended to achieve QIT between any two selected NVCEs in a large hybrid system by adjusting system parameters, which is important in large scale quantum information processing.
\end{abstract}
 \maketitle
\section{introduction}
Quantum information transfer (QIT) has many applications in communication
science \cite{a0}. There exist physical systems for realizing QIT, such as,
cavity quantum electrodynamics (QED) \cite{a01, a02, a021, a022}, linear
optics devices \cite{a03}, and superconducting qubits \cite{a04, a05, a06,
a07, a08}, \emph{etc}. In addition, a nitrogen-vacancy center in diamond has
been recently considered as one of the most promising candidates for quantum
information processing, due to its relatively long coherence time and the
possibility of coherent manipulation at room temperature \cite{b}. For
instances, the electron spin relaxation time $T_{1}=6$ms \cite{c} and
isotopically pure diamond sample dephasing time $T_{2}=2$ms \cite{d} have
been reported, coherent oscillations in a single electron spin have been
observed \cite{d1}, and coherent time of a nitrogen-vacancy center has been
improved very much in the recent years and could reach $1$ second \cite{d11}%
. On the other hand, hybrid solid-state devices have attracted tremendous
attentions (see \cite{a} and references therein). Theoretically, the
physical systems, composed of spin ensembles and superconducting qubits
fabricated in a TLR (transmission line resonator), have been proposed \cite%
{a1, a2, a3, a33}. Experimentally, a quantum circuit consisting of a
superconducting qubit and a nitrogen-vacancy center ensemble (NVCE) has been
implemented in Ref. \cite{a4}; and a quantum SWAP gate has been realized in
this circuit, by employing the strong coupling between a superconducting
qubit and a NVCE \cite{a4}. In addition, Marcos \emph{et al.} \cite{a5} have
proposed a hybrid system, in which the direct coupling between a
superconducting flux qubit and a NVCE is much stronger than that between a
NVCE and a TLR. For the work on the coupling between a NVCE and a TLR, see
Refs. \cite{d3, d4}. Experimentally, the strong coupling between a
superconducting flux qubit and a NVCE has been demonstrated \cite{a6}.
Moreover, by using the strong coupling, the QIT between a flux qubit and a
NVCE has been performed in experiment \cite{a7}. Then, the strong coupling between
 a NVCE and a TLR via a flux qubit used as a data bus was proposed in Ref. \cite{a8}. These results provide a
platform for using NVCEs as quantum memories, which are essential in quantum
information processing.

Motivated by the recent works on the coupling between \emph{LC} circuits and
flux qubits \cite{a9, a10, a11, a12}, and the strong coupling hybrid solid quantum system \cite{a5, a6, a7, a8}, as well as the QIT
 with the solid quantum system \cite{a80, a81, a82}, we
will propose an architecture for scalable QIT among NVCEs. In this
architecture, a \emph{LC} circuit is used to induce the necessary
interaction between flux qubits, each magnetically coupling to a NVCE. We
explicitly show that for resonant interaction and large detuning cases,
high-fidelity QIT between the two spatially-separated NVCEs can be
implemented by solving Schr\"{o}dinger equations. Moreover, this
architecture can be extended to scale up multiple flux qubits and NVCEs by
using a single \emph{LC} circuit, and the QIT between any two selected NVCEs
can be achieved in this large hybrid system. To the best of our knowledge,
how to realize QIT between NVCEs in this architecture has not been proposed
yet. Note that, Refs. \cite{a13, a14} reported the QIT between two ensembles
which are trapped in spatially separated cavities, respectively. But, the
fidelity of the QIT was not calculated and the dissipation of the system was
not considered in Refs. \cite{a13, a14}.

\section{Model}
\begin{figure}
  \centering
  \includegraphics[scale=0.4]{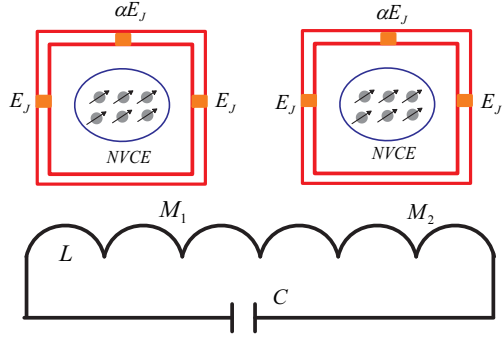}\\
  \caption{(Color online) Quantum information transfer circuit. Two flux qubits are coupled to a \emph{LC} circuit by their mutual inductances $M_{j}$ (j=1,2). Each flux qubit consists of three Josephson junctions, i.e., two large junctions with the same coupling energy $E_{J}$ and a small
junction with the coupling energy $\alpha E_{J} ~(0.5<\alpha<1)$.
  The separation between the two qubits is assumed to be much larger than the linear dimension of
each qubit,  such that the direct interaction between the two flux qubits is negligible. Here, the NVCEs play a role of memory units, which are used to store quantum information for a long time.}\label{1}
\end{figure}
We propose a QIT hybrid circuit, as shown in Fig. (\ref{1}), which consists of a \emph{LC} circuit acting as a
data bus to induce coupling between two flux qubits. Each flux qubit couples to a NVCE by a magnetic field. Each NVCE is an information memory unit.
The electronic ground state of a single nitrogen-vacancy center (NVC) has a spin $\mathbb{S}=1$, with the levels $m_{s}=0$ and $m_{s}=\pm1$ separated by zero-field splitting $D$. For a NVC, the Hamiltonian can be described by (assuming $\hbar=1$) \cite{e0, e}
\begin{eqnarray}
H_{NVC}=D\mathbb{S}^{2}_{z}+E(\mathbb{S}^{2}_{x}-\mathbb{S}^{2}_{y})+g_{e}\mu_{B}\vec{B}\cdot\vec{\mathbb{S}},
\end{eqnarray}
where zero field splitting $D=2.88$ GHz, $\vec{\mathbb{S}}=\{\mathbb{S}_{x}, \mathbb{S}_{y}, \mathbb{S}_{z}\}$ is a usual Pauli spin-1 operator, $E$ is the strain-induced splitting coefficient, $B$ is the applied magnetic field, $g_{e}$ is the Lande factor, and $\mu_{B}$ is the Bohr magneton. When the static magnetic field $\vec{B}$ is applied along the crystalline axis of the diamond, the degeneracy of levels $|m_{s}=\pm1\rangle$ can be removed. The quantum information is encoded in sublevels $|m_{s}=0\rangle\equiv|0\rangle$ and $|m_{s}=-1\rangle\equiv|1\rangle$ serving as two logic states of a
qubit. For a NVCE with NVCs ($1,2,\ldots N$), the ground state is defined as $|g\rangle=|0_{1}\cdots0_{k}\cdots0_{N}\rangle$ while the excited state is defined as $|e\rangle=S^{+}|g\rangle=(1/\sqrt{N})\sum_{k=1}^{N}|0_{1}\cdots1_{k}\cdots0_{N}\rangle$ with operator $S^{+}=(S^{-})^{\dag}=(1/\sqrt{N})\sum_{k}^{N}|1\rangle_{k}\langle0|$, where the subscript \emph{k} represents the \emph{k}-th NVC. Thus, the Hamiltonian of a NVCE is written as \cite{a3} $H_{NVCE}=\frac{1}{2}\Omega S_{z}$, where $\Omega=D-g_{e}\mu_{B}B_{z}$ is the energy gap between the
ground state $|g\rangle$ and the excited state $|e\rangle$, with the operator $S_{z}=|e\rangle\langle e|-|g\rangle\langle g|$.

The Hamiltonian of a flux qubit is described as a two-level system \cite{f0, f}
\begin{eqnarray}
H_{q}=\frac{1}{2}(\varepsilon\sigma_{z}+\Delta\sigma_{x}),
\end{eqnarray}
where $\varepsilon(\Phi)=2I_{p}[\Phi-(1/2+n)\Phi_{0}]$ is the energy spacing of the two classical current
states, $I_{p}$ is persistent current of the
flux qubit, $\Phi_{0}=h/2e$ is the magnetic-flux
quantum, $\Phi$ is the external magnetic flux applied to the qubit loop, $\Delta$ is the energy gap between the two energy levels of the qubit at the degeneracy
point, and Pauli matrices $\sigma_{z}=|b\rangle\langle b|-|a\rangle\langle a|$ and $\sigma_{x}=|b\rangle\langle a|+|a\rangle\langle b|$ are defined in terms of the classical current,
with $|a\rangle=|\circlearrowright\rangle$ and $|b\rangle=|\circlearrowleft\rangle$ denoting the states with clockwise and counterclockwise
currents in the qubit loop. In terms of the eigenbasis of the flux qubit, the Hamiltonian (2) can be rewritten as $H_{q}=\frac{1}{2}\omega_{q}\sigma_{z}$, with
$\hbar\omega_{q}=\sqrt{\varepsilon^{2}+\Delta^{2}}$ being the energy level separation of the flux qubit.

As long as the distance between the two flux qubits is large, the direct interaction between the
two flux qubits is negligible. For a system in Fig. 1, the total Hamiltonian is given by
\begin{eqnarray}
H&=&\omega a^{\dag}a+\sum_{j=1}^{2}[\frac{1}{2}\omega^{j}_{q}\sigma^{j}_{z}+g_{j}(a\sigma_{j}^{+}+a^{\dag}\sigma_{j}^{-})\nonumber\\
&&+\frac{1}{2}\Omega_{j} S^{j}_{z}+J_{j}(S_{j}^{+}\sigma_{j}^{-}+S_{j}^{-}\sigma_{j}^{+})],
\end{eqnarray}
where the first term is the free Hamiltonian of the \emph{LC} circuit with the resonance frequency $\omega=1/\sqrt{LC}$ and the plasmon annihilation (creation) operator $a~(a^{\dag})$ \cite{a11}, the third term represents the interaction between the \emph{LC} circuit and the flux qubits with the coupling constant $g_{j}=M_{j}I_{p}\sqrt{\omega/2L}$ \cite{a11} and the operator $\sigma^{\dag}_{j}=(\sigma^{-}_{j})^{\dag}=|1\rangle_{j}\langle0|$,
the last term indicates the coupling between NVCEs and flux qubits with the coupling strength $J_{j}$ \cite{a5}.

\section{Quantum information transfer}
In this section, we discuss how to realize QIT between spatially-separated two NVCEs for both resonant interaction and large detuning cases. By solving Schr\"{o}dinger equations, we find that high-fidelity QIT can be implemented at some moment, as shown below.

For simplicity, we use \emph{NE} to represent NVCE in each equation below, but still use NVCE in the word text.
\subsection{Resonant interaction case}
In the interaction picture, the Hamiltonian of the total system for the resonant interaction case (i.e. $\omega=\omega_{q}^{j}=\Omega_{j}$) can be written as follows
\begin{eqnarray}
H_{I}=\sum_{j=1}^{2}[g_{j}(a\sigma_{j}^{+}+a^{\dag}\sigma_{j}^{-})+J_{j}(S_{j}^{+}\sigma_{j}^{-}+S_{j}^{-}\sigma_{j}^{+})]. \label{a}
\end{eqnarray}
The QIT from the left NVCE (i.e., $NE_{1}$) to the right one (i.e., $NE_{2}$) is described by the  formula $(\alpha|g\rangle+\beta|e\rangle)_{NE_{1}}|0\rangle_{1}|0\rangle_{2}|0\rangle_{L}|g\rangle_{NE_{2}}\rightarrow
|g\rangle_{NE_{1}}|0\rangle_{1}|0\rangle_{2}|0\rangle_{L}(\alpha|g\rangle+\beta|e\rangle)_{NE_{2}}$, where the subscripts $1,~ 2$, and $L$ represent the left flux qubit, the right flux qubit, and \emph{LC} circuit, respectively;  $\alpha$ and $\beta$ are the normalized complex numbers.
When the initial state of the system is $|\Psi(0)\rangle=|e\rangle_{NE_{1}}|0\rangle_{1}|0\rangle_{2}|0\rangle_{L}|g\rangle_{NE_{2}}$, the system state evolves in the subspace $\{|\varphi_{1}\rangle, |\varphi_{2}\rangle, |\varphi_{3}\rangle, |\varphi_{4}\rangle, |\varphi_{5}\rangle\}$ with
\begin{subequations}
\begin{align}
|\varphi_{1}\rangle&=|e\rangle_{NE_{1}}|0\rangle_{1}|0\rangle_{2}|0\rangle_{L}|g\rangle_{NE_{2}}, \\
|\varphi_{2}\rangle&=|g\rangle_{NE_{1}}|1\rangle_{1}|0\rangle_{2}|0\rangle_{L}|g\rangle_{NE_{2}}, \\
|\varphi_{3}\rangle&=|g\rangle_{NE_{1}}|0\rangle_{1}|1\rangle_{2}|0\rangle_{L}|g\rangle_{NE_{2}}, \\
|\varphi_{4}\rangle&=|g\rangle_{NE_{1}}|0\rangle_{1}|0\rangle_{2}|1\rangle_{L}|g\rangle_{NE_{2}}, \\
|\varphi_{5}\rangle&=|g\rangle_{NE_{1}}|0\rangle_{1}|0\rangle_{2}|0\rangle_{L}|e\rangle_{NE_{2}},
\end{align}
\end{subequations}
where $|g\rangle_{NE_{j}}$ and $|e\rangle_{NE_{j}} ~(j=1,2)$ are, respectively, the ground state and the symmetric Dicke excitation state of the \emph{j}-th NVCE, $|0\rangle_{j}$ ($|1\rangle_{j}$) is the ground (excited) state of the \emph{j}-th flux qubit; $|0\rangle_{L}$ ($|1\rangle_{L}$) is the ground (single-excited) state of the \emph{LC} circuit. At any instant, the quantum state of the system is described by
\begin{eqnarray}
|\Psi(t)\rangle=\sum_{i=1}^{5}C_{i}(t)|\varphi_{i}\rangle,
\end{eqnarray}
where the normalized coefficients satisfy $\sum_{i=1}^{5}|C_{i}(t)|^{2}=1$. Suppose the two flux qubits equally couple to the \emph{LC} circuit ($g_{1}=g_{2}=g$) and equally couple with their NVCEs ($J_{1}=J_{2}=J$). In this case, for the initial conditions $C_{1}(0)=1$ and $C_{2}(0)=C_{3}(0)=C_{4}(0)=C_{5}(0)=0$, we can easily get the expression of the time-dependent coefficients,
\begin{subequations}
\begin{align}
C_{1}(t)&=\frac{g^{2}}{J^{2}+2g^{2}}+\frac{1}{2}\cos Jt+\frac{J^{2}}{2(J^{2}+2g^{2})}\cos\sqrt{J^{2}+2g^{2}}t, \\
C_{2}(t)&=-i\frac{1}{2}\sin Jt-i\frac{J}{2\sqrt{J^{2}+2g^{2}}}\sin\sqrt{J^{2}+2g^{2}}t, \\
C_{3}(t)&=i\frac{1}{2}\sin Jt-i\frac{J}{2\sqrt{J^{2}+2g^{2}}}\sin\sqrt{J^{2}+2g^{2}}t, \\
C_{4}(t)&=-\frac{Jg}{J^{2}+2g^{2}}+\frac{Jg}{J^{2}+2g^{2}}\cos\sqrt{J^{2}+2g^{2}}t, \\
C_{5}(t)&=\frac{g^{2}}{J^{2}+2g^{2}}-\frac{1}{2}\cos Jt+\frac{J^{2}}{2(J^{2}+2g^{2})}\cos\sqrt{J^{2}+2g^{2}}t.
\end{align}
\end{subequations}

The quantum state $|g\rangle_{NE_{1}}|0\rangle_{c1}|0\rangle_{c2}|0\rangle_{L}|g\rangle_{NE_{2}}$ remains unchanged under the Hamiltonian (\ref{a}). Thus, when the quantum state $|\Psi(t)\rangle$ collapses into $|\varphi_{5}\rangle$, the quantum information is transferred from the left NVCE (i.e., $NE_{1}$) to the right one (i.e., $NE_{2}$). Hence, the populations of quantum states $|\varphi_{1}\rangle$ and $|\varphi_{5}\rangle$ are important measure for the QIT. Our proposal includes two coupling mechanisms: the magnetical coupling $J$ between the flux qubits and the NVCEs, the mutual-inductance coupling $g$ between the flux qubits and the \emph{LC} circuit. Next, according to the relation between coupling strengths $g$ and $J$, we will analyze the  populations of quantum states $|\varphi_{1}\rangle$ and $|\varphi_{5}\rangle$.

Case (i) (the case for the equilibrium coupling $g=J$): the coefficients of quantum state $|\varphi_{1}\rangle$ and $|\varphi_{5}\rangle$ can be written as $C_{1}(t)=1/3+1/2\cos Jt+1/6\cos\sqrt{3}Jt$ and $C_{5}(t)=1/3-1/2\cos Jt+1/6\cos\sqrt{3}Jt$, respectively. In Fig. 2(a), we plot the population change with $Jt$. Obviously, $|C_{5}(t)|^{2}$ can reach the maximum at some moment. This means that QIT between spatially-separated two NVCEs can be perfectly realized.

Case (ii) (the case for the strong
magnetic coupling $J\gg g$): if $J\gg g$, $C_{5}(t)$ tends to zero. This result shows that the QIT can not be realized in our system. We plot $|C_{1}(t)|^{2}$ and $|C_{5}(t)|^{2}$ for a coupling strength $J=10g$ in Fig. 2(b), which shows that the QIT between spatially-separated two NVCEs can be realized, but it takes a longer time.

Case (iii) (the case for the strong mutual inductance
coupling $J\ll g$): if $J\ll g$, the expressions of $C_{1}(t)$ and $C_{5}(t)$ are reduced to $C_{1}(t)=1/2+1/2\cos Jt$ and $C_{5}(t)=1/2-1/2\cos Jt$, respectively. When $Jt=(2k+1)\pi ~(k=0,1,2\ldots)$, one has $C_{1}(t)=0$, but $C_{5}(t)=1$, which means that the information has been transferred from the left NVCE to the right one. We have plotted Fig. 2(c) to show how $|C_{1}(t)|^{2}$ and $|C_{5}(t)|^{2}$ change with time $t$ for a coupling strength $J=0.1g$. Fig. 2(c) shows that the QIT between spatially-separated two NVCEs can be implemented.

The recent experiments have reported that the effective coupling strength between a flux qubit and a NVCE (containing $N\sim3.1\times10^{7}$ NVCs) can reach $J\sim70$MHz \cite{a6}, and the coupling strength between a flux qubit and a \emph{LC} circuit can reach $g=220$MHz \cite{a9}. Hence, the condition $J\ll g$ for Case (iii) can be well satisfied.

However, in real physical systems the coupling strengths between flux qubits and LC circuit (or NVCEs) are not the same. The expressions of the time-dependent coefficients given in Eqs. (7) become rather long and complicated for the unbalanced coupling case. Here, we only numerically simulate the population change of quantum states with time, as shown in Fig. \ref{22}. It can be seen from Fig. \ref{22} that the perfect QIT between spatially-separated two NVCEs can also be realized except for the unbalanced strong magnetic coupling case.

\begin{figure}
  \centering
  \includegraphics[scale=0.45]{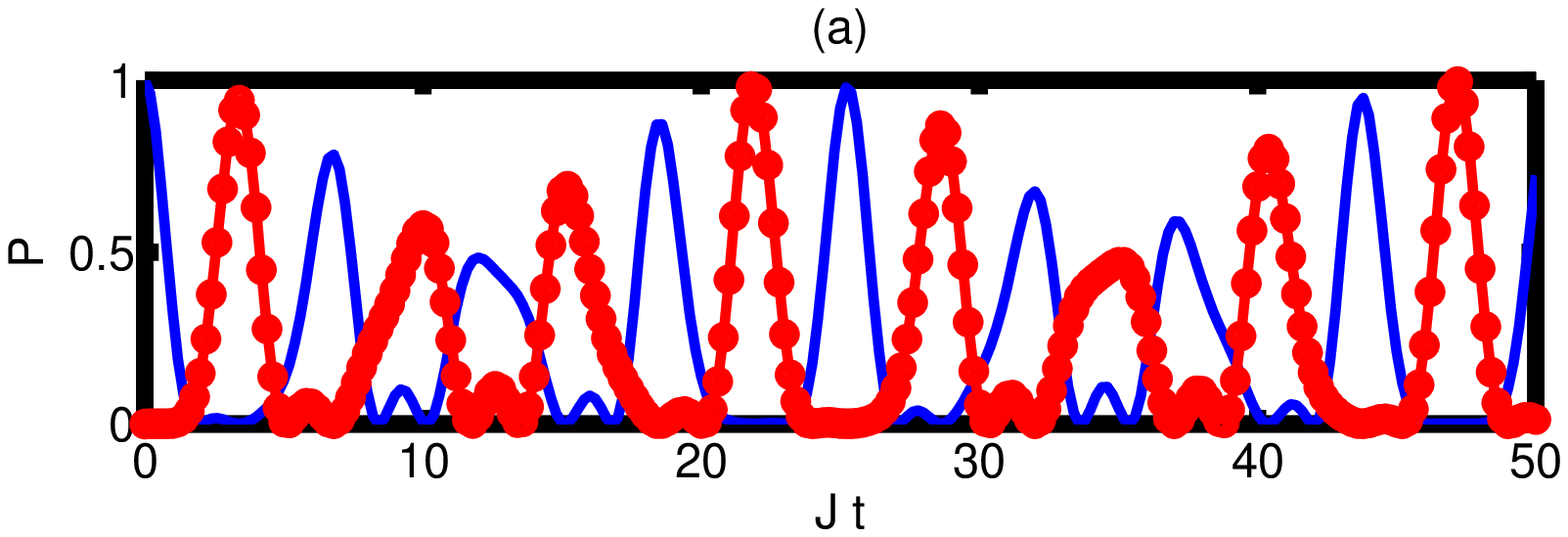}\\
  \includegraphics[scale=0.45]{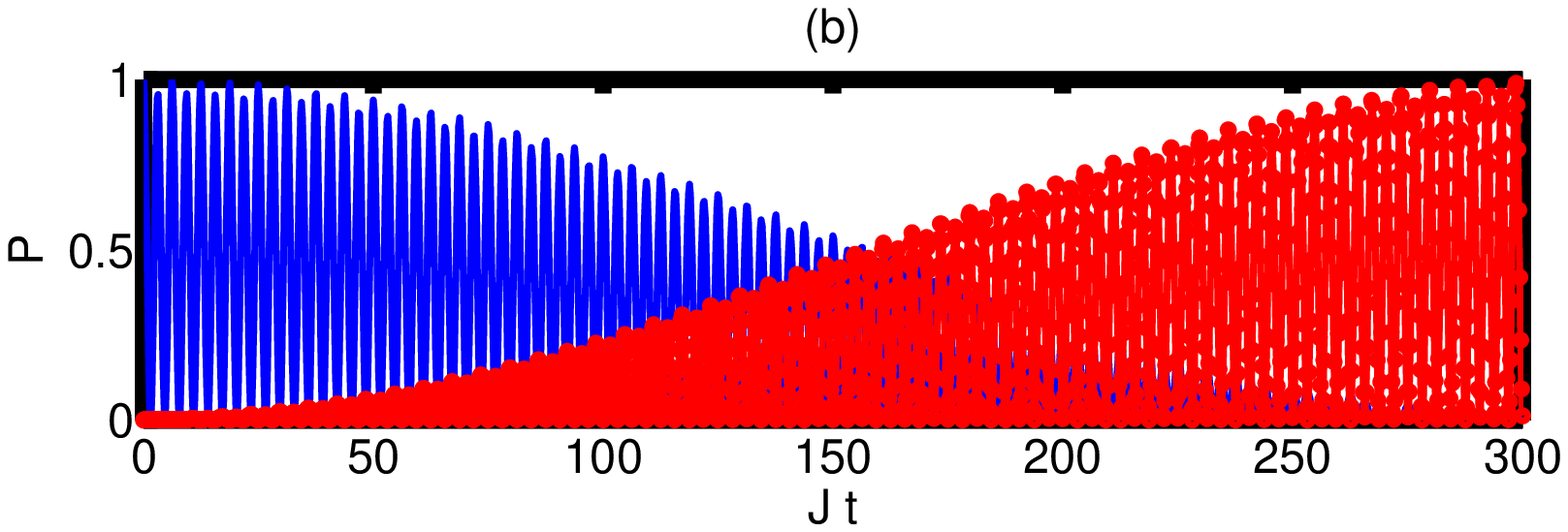}\\
  \includegraphics[scale=0.45]{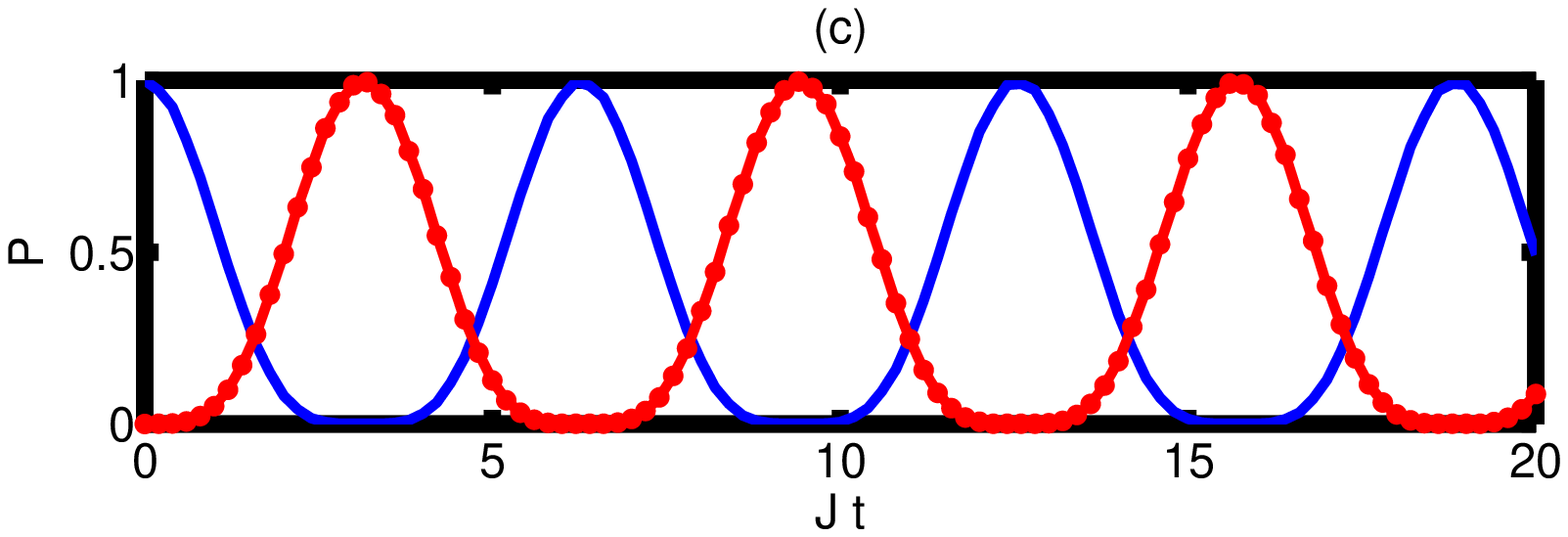}\\
  \caption{(Color online) Solid-blue lines represent the population $|C_{1}|^2$ of the state $|\varphi_{1}\rangle$, while solid-dot-red lines indicate the population $|C_{5}|^{2}$ of the state $|\varphi_{5}\rangle$, (a) the equilibrium coupling $J=g$, (b) the strong magnetic coupling $J=10g$, (c) the strong mutual inductance coupling $J=0.1g$.}\label{2}
\end{figure}

\begin{figure}
  \centering
  \includegraphics[scale=0.4]{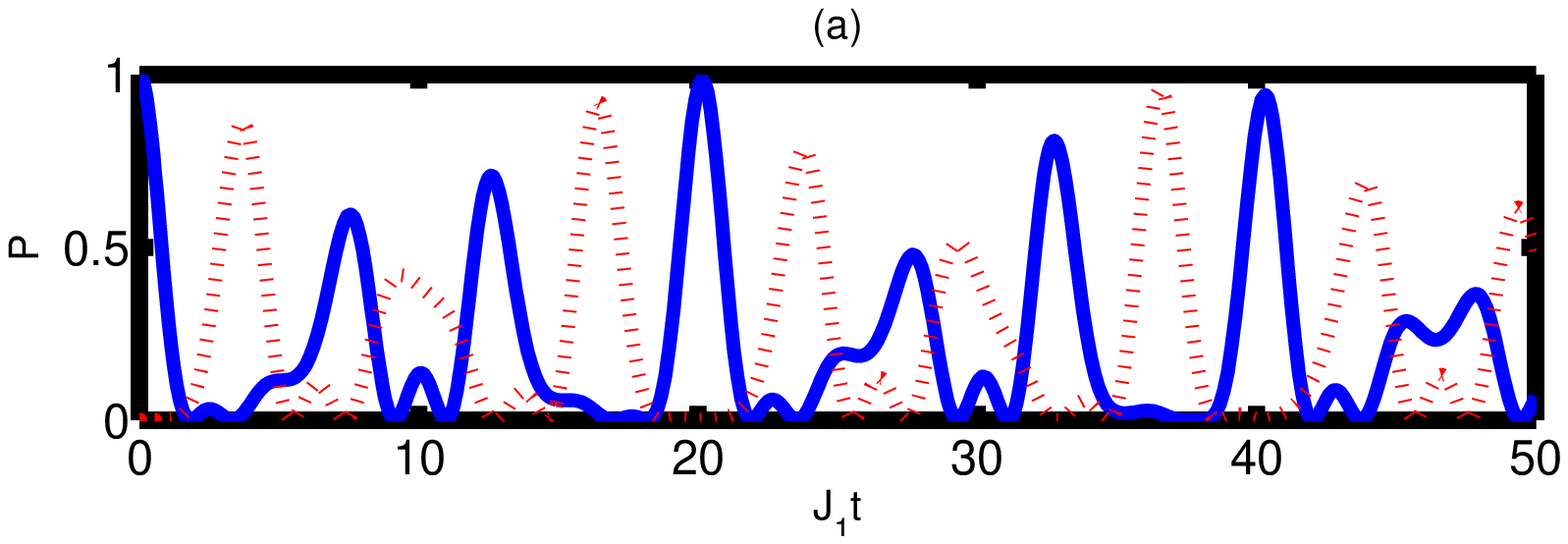}\\
  \includegraphics[scale=0.4]{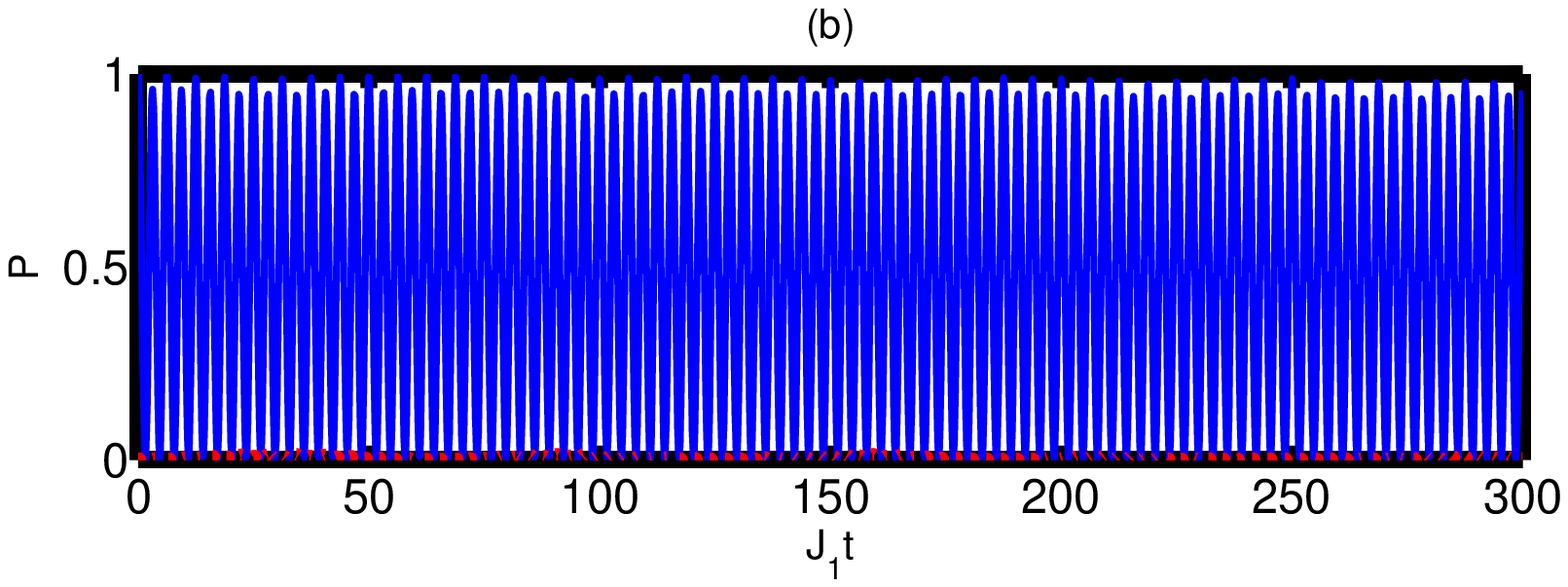}\\
  \includegraphics[scale=0.4]{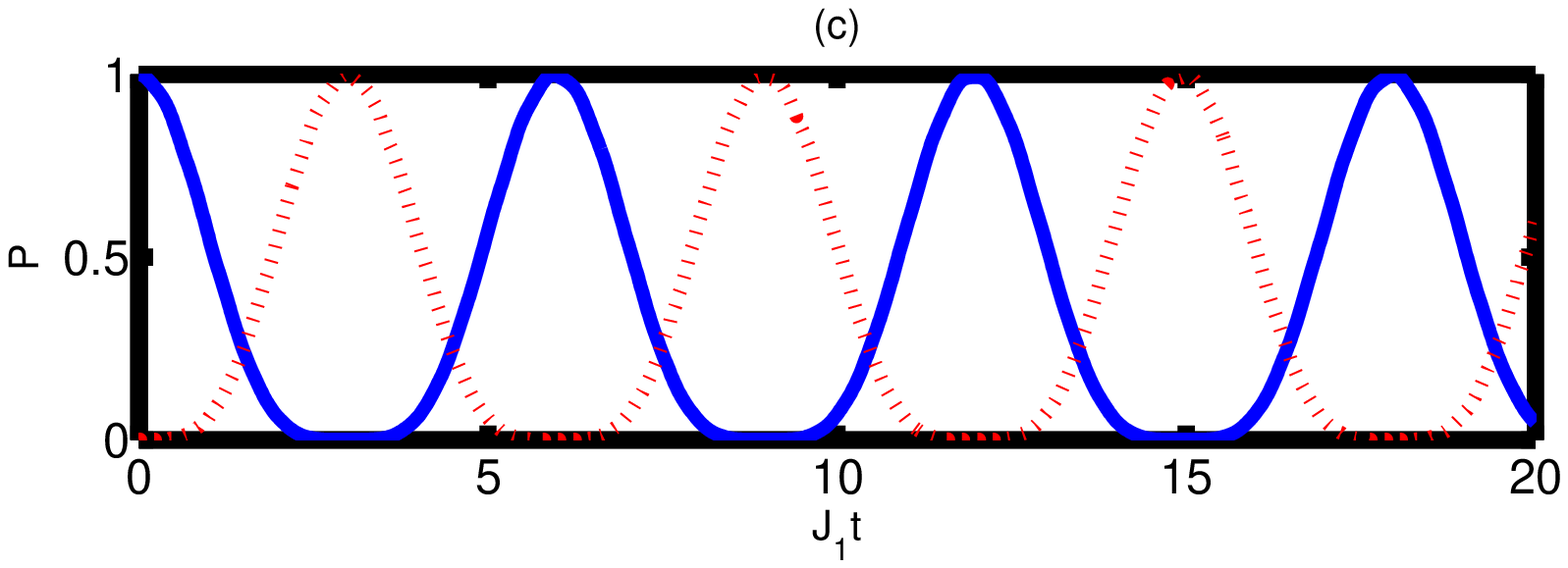}\\
  \caption{(Color online) Solid-blue lines represent the population $|C_{1}|^2$ of the state $|\varphi_{1}\rangle$, while dash lines indicate the population $|C_{5}|^{2}$ of the state $|\varphi_{5}\rangle$, (a) the unbalanced equilibrium coupling $g_{1}=0.9J_{1},~g_{2}=0.9J_{2},~J_{2}=0.9J_{1}$, (b) the unbalanced strong magnetic coupling $J_{1}=10g_{1},~J_{2}=10g_{2},~g_{1}=0.9g_{2}$, (c) the unbalanced strong mutual inductance coupling $J_{1}=0.1g_{1},~J_{2}=0.1g_{2}, g_{1}=0.9g_{2}$.}\label{22}
\end{figure}
\subsection{Large detuning case}
In this section, we will show how to realize QIT between two NVCEs within a large detuning regime. We will only consider the large detuning between the \emph{LC} circuit and the flux qubits, but still apply the resonance interaction between the flux qubits and the NVCEs. In the interaction picture, the Hamiltonian for the system (shown in Fig. \ref{1}) is
\begin{eqnarray}
H_{I}=\sum_{j=1}^{2}[g_{j}(a\sigma_{j}^{+}e^{i\delta_{j} t}+a^{\dag}\sigma_{j}^{-}e^{-i\delta_{j} t})+J_{j}(S_{j}^{+}\sigma_{j}^{-}+S_{j}^{-}\sigma_{j}^{+})],
\end{eqnarray}
where $\delta_{j}=\omega_{q}^{j}-\omega$ is the detuning between the transition frequency of the \emph{j}-th flux qubit and the frequency of the \emph{LC} circuit. In the large detuning case $\delta_{j}\gg g_{j}$, there is no energy exchange between the flux qubits and the \emph{LC} circuit. Accordingly, there is no energy exchange between each NVCE and the \emph{LC} circuit. We consider that two identical flux qubits simultaneously
interact with the \emph{LC} circuit and assume that the
\emph{LC} circuit is initially in the vacuum state. Then, the effective
Hamiltonian is given by \cite{a01}
\begin{eqnarray}
H_{eff}&=&\sum_{j=1}^{2}\lambda_{j}[(|1\rangle_{j}\langle1|+\sigma_{1}^{\dag}\sigma_{2}^{-}+\sigma_{1}^{-}\sigma_{2}^{\dag})\nonumber\\
&&+J_{j}(S_{j}^{+}\sigma_{j}^{-}+S_{j}^{-}\sigma_{j}^{+})], \label{b}
\end{eqnarray}
where $\lambda_{j}=g_{j}^{2}/\delta_{j}$. The first term describes the \emph{LC}-induced energy stark shift; the second and third terms represent the dipole coupling between the
two flux qubits, induced by the \emph{LC} circuit; and the last two terms represent the interaction between the NVCEs and the flux qubits. The virtual excitation
of the \emph{LC} circuit avoids the population loss of the data bus.

We assume that quantum information is initially encoded in the left NVCE (i.e., $NE_{1}$). Because the state $|g\rangle_{NE_{1}}|0\rangle_{1}|0\rangle_{2}|g\rangle_{NE_{2}}$ remains unchanged under the Hamiltonian (\ref{b}), we only need to care about the evolution of the state $|\Psi(0)\rangle=|e\rangle_{NE_{1}}|0\rangle_{1}|0\rangle_{2}|g\rangle_{NE_{2}}$. The system state evolves within
the subspace, formed by the following states
\begin{subequations}
\begin{align}
|\phi_{1}\rangle&=|e\rangle_{NE_{1}}|0\rangle_{1}|0\rangle_{2}|g\rangle_{NE_{2}},  \\
|\phi_{2}\rangle&=|g\rangle_{NE_{1}}|1\rangle_{1}|0\rangle_{2}|g\rangle_{NE_{2}},  \\
|\phi_{3}\rangle&=|g\rangle_{NE_{1}}|0\rangle_{1}|1\rangle_{2}|g\rangle_{NE_{2}},  \\
|\phi_{4}\rangle&=|g\rangle_{NE_{1}}|0\rangle_{1}|0\rangle_{2}|e\rangle_{NE_{2}}.
\end{align}
\end{subequations}
The quantum state of the system at any time is expressed as
\begin{eqnarray}
|\Psi(t)\rangle=\sum_{i=1}^{4}D_{i}(t)|\phi_{i}\rangle,
\end{eqnarray}
where the normalized coefficients satisfy $\sum_{i=1}^{4}|D_{i}(t)|^{2}=1$. For the initial condition $D_{1}(0)=1$ and $D_{2}(0)=D_{3}(0)=D_{4}(0)=0$, and for the identical coupling strengths between the flux qubits and the NVCEs (i.e. $J_{1}=J_{2}=J$), and the two flux qubits equally to the \emph{LC} circuit ($\lambda_{1}=\lambda_{2}=\lambda$), we can easily obtain the following time-dependent coefficients
\begin{subequations}
\begin{align}
D_{1}(t)&=\frac{J^{2}}{4(\kappa^{2}+\lambda\kappa)}e^{-i(\lambda+\kappa)t}
+\frac{J^2}{4(\kappa^{2}-\lambda\kappa)}e^{-i(\lambda-\kappa)t}+\frac{1}{2}\cos Jt, \\
D_{2}(t)&=\frac{J(\lambda+\kappa)}{4(\kappa^{2}+\lambda\kappa)}e^{-i(\lambda+\kappa)t}
+\frac{J(\lambda-\kappa)}{4(\kappa^{2}-\lambda\kappa)}e^{-i(\lambda-\kappa)t}
-i\frac{1}{2}\sin Jt, \\
D_{3}(t)&=\frac{J(\lambda+\kappa)}{4(\kappa^{2}+\lambda\kappa)}e^{-i(\lambda+\kappa)t}
+\frac{J(\lambda-\kappa)}{4(\kappa^{2}-\lambda\kappa)}e^{-i(\lambda-\kappa)t}
+i\frac{1}{2}\sin Jt, \\
D_{4}(t)&=\frac{J^{2}}{4(\kappa^{2}+\lambda\kappa)}e^{-i(\lambda+\kappa)t}
+\frac{J^2}{4(\kappa^{2}-\lambda\kappa)}e^{-i(\lambda-\kappa)t}-\frac{1}{2}\cos Jt,
\end{align}
\end{subequations}
with the parameter $\kappa=\sqrt{\lambda^2+J^2}$. The information exchange between the two NVCEs can be characterized by the population change of the quantum states $|\phi_{1}\rangle$ and $|\phi_{4}\rangle$. Following the resonant interaction case, we now discuss the relation between the $|D_{1}(t)|^{2}$ and $|D_{4}(t)|^{2}$ for different dipole-dipole coupling strength $\lambda$ and magnetical coupling strength $J$. For the equilibrium coupling $\lambda=J$, we plot the population evolution in Fig. 4(a). The perfect QIT can be achieved at some moment. Comparing Fig. 4(a) with Fig. 2(a), one can see that the time required for QIT is shorter than that for the resonant interaction case. Fig. 4(b) shows that for the strong magnetic coupling $J=10\lambda$, the QIT can be realized and the required time is reduced by one order of magnitude, compared with Fig. 2(b). For the stronge dipole-dipole coupling $J=0.1\lambda$, the QIT can also be realized, as shown in Fig. 4(c). But the successful probability of the QIT decreases as the time increases. For the unbalanced coupling case, we only numerically simulate the changing of the $|D_{1}(t)|^{2}$ and $|D_{4}(t)|^{2}$ with $J_{1}t$ as shown in Fig. \ref{33}, which shows that the QIT between two NVCEs can also be implemented.
\begin{figure}
  \centering
  \includegraphics[scale=0.45]{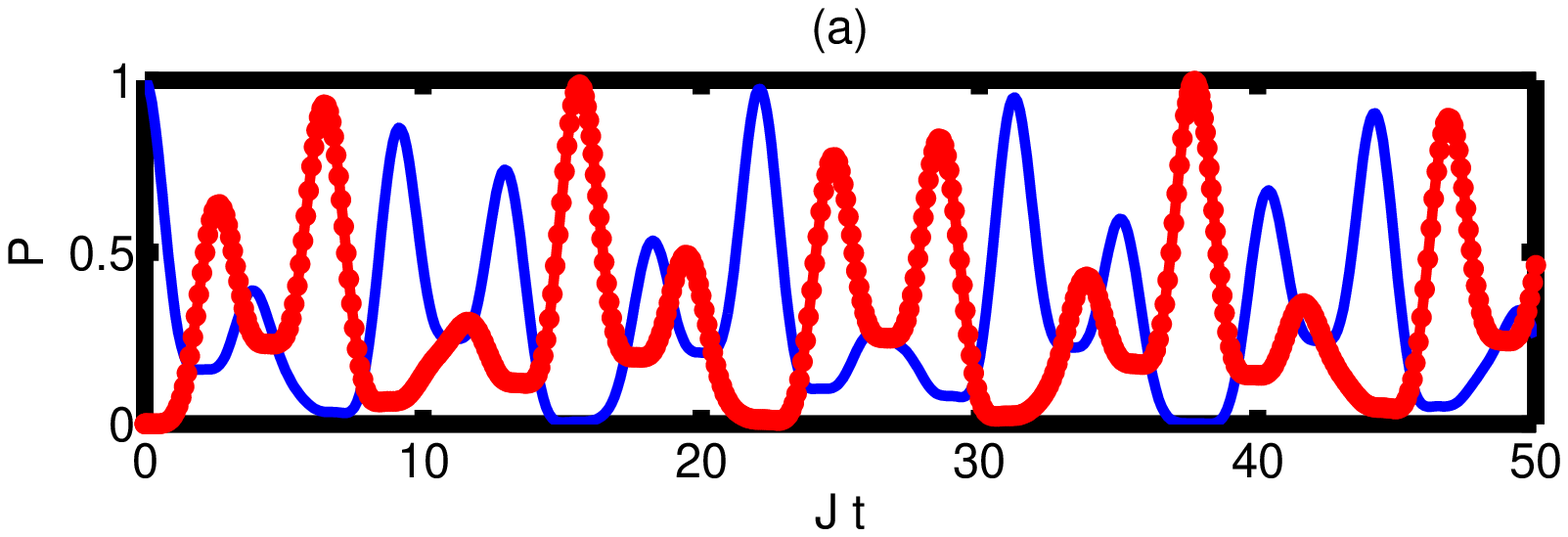}\\
  \includegraphics[scale=0.45]{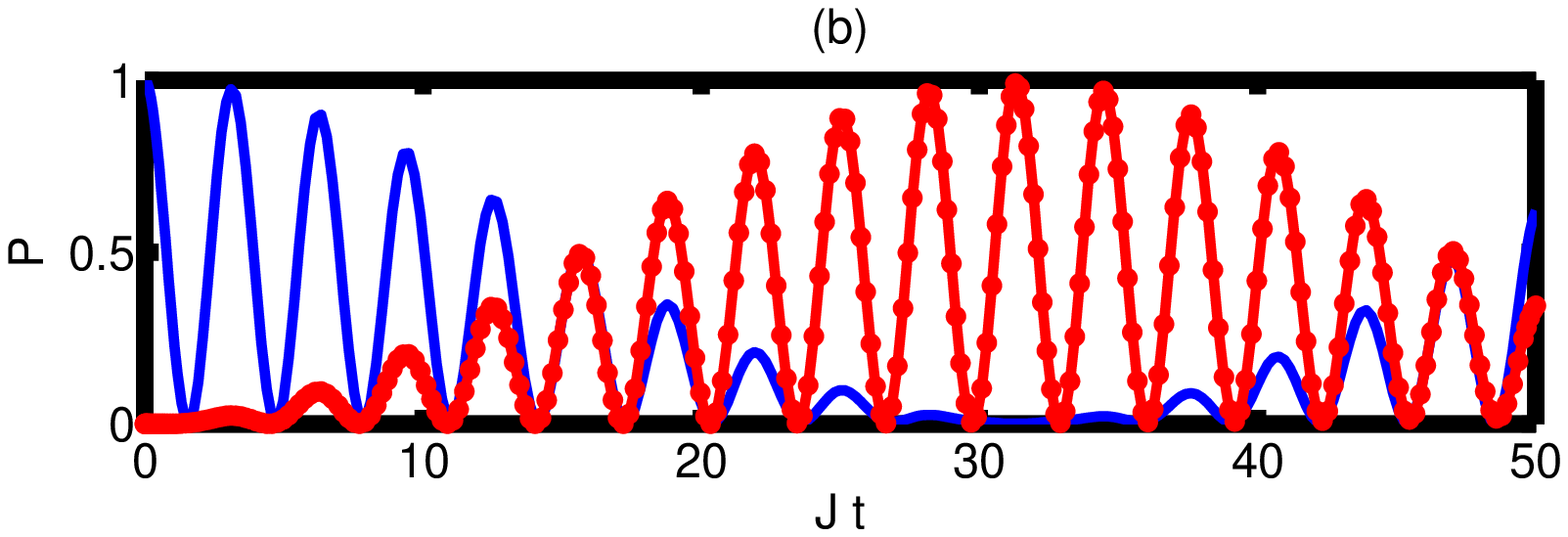}\\
  \includegraphics[scale=0.45]{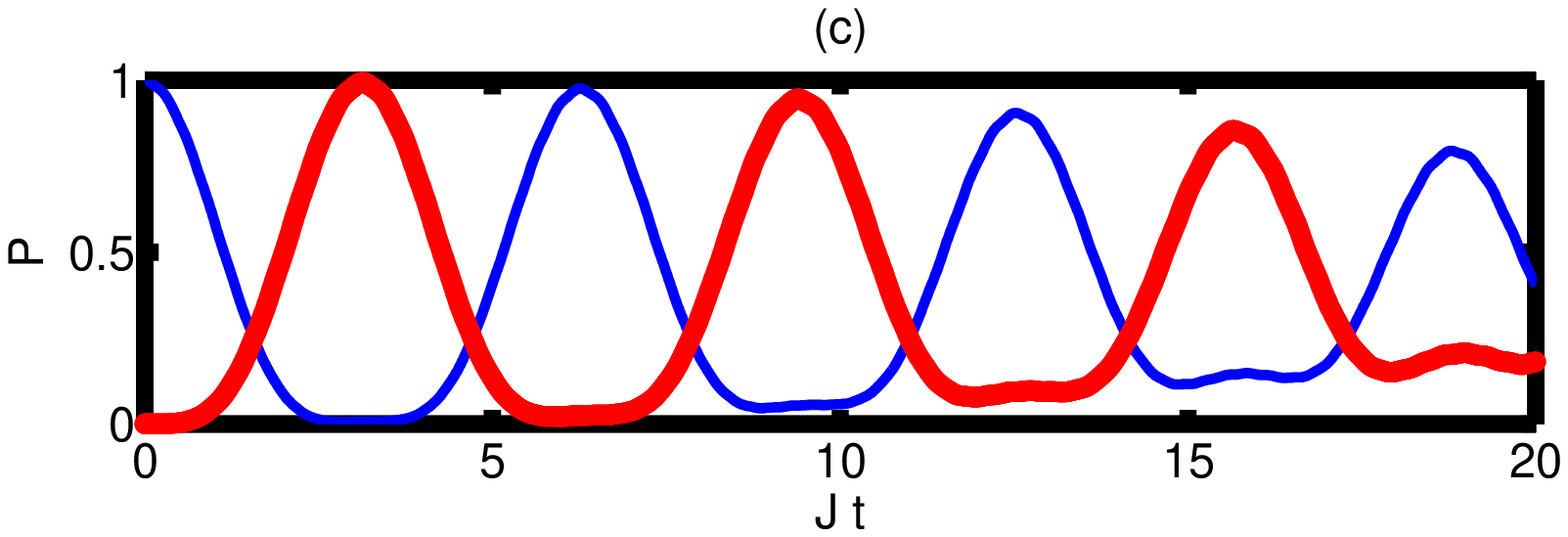}\\
  \caption{(Color online) The time evolution of the populations of the states $|\phi_{1}\rangle$ and $|\phi_{4}\rangle$ for the different coupling mechanisms (a) $J=\lambda$, (b) $J=10\lambda$, and (c) $J=0.1\lambda$. Solid-blue lines represent the population $|D_{1}(t)|^{2}$ of the state $|\phi_{1}\rangle$, while solid-dot-red lines indicate the the population $|D_{4}(t)|^{2}$ of the state $|\phi_{4}\rangle$.}\label{3}
\end{figure}
\begin{figure}
  \centering
  \includegraphics[scale=0.45]{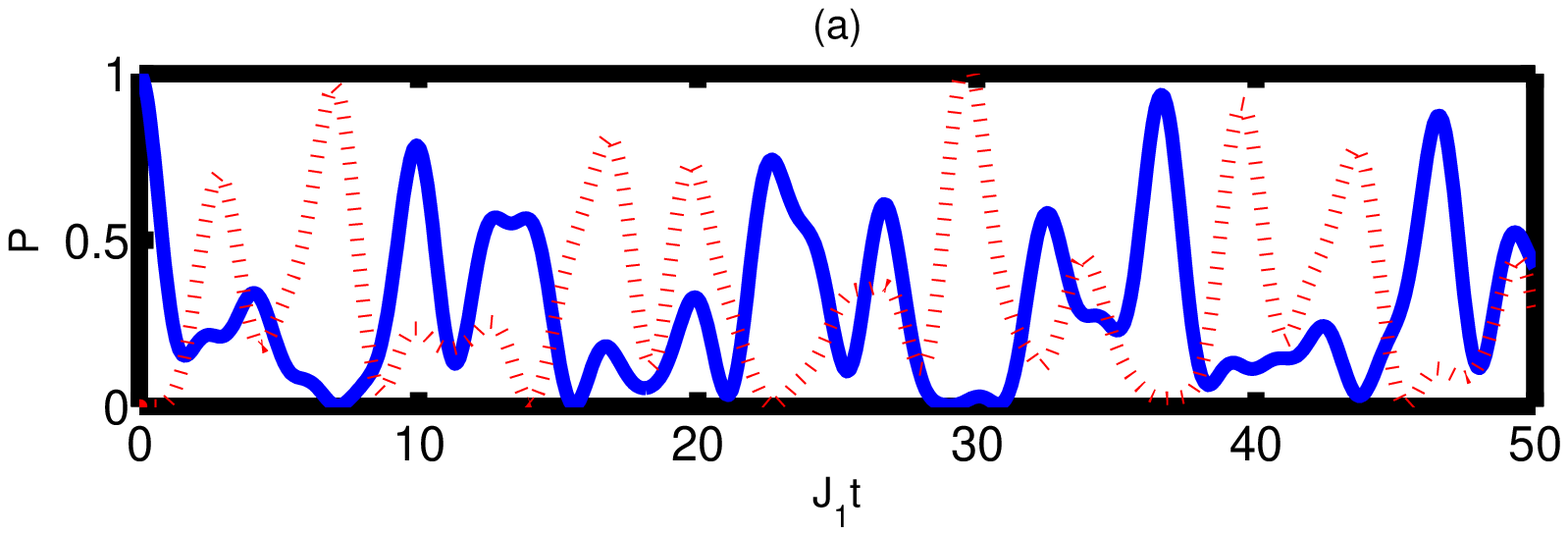}\\
  \includegraphics[scale=0.45]{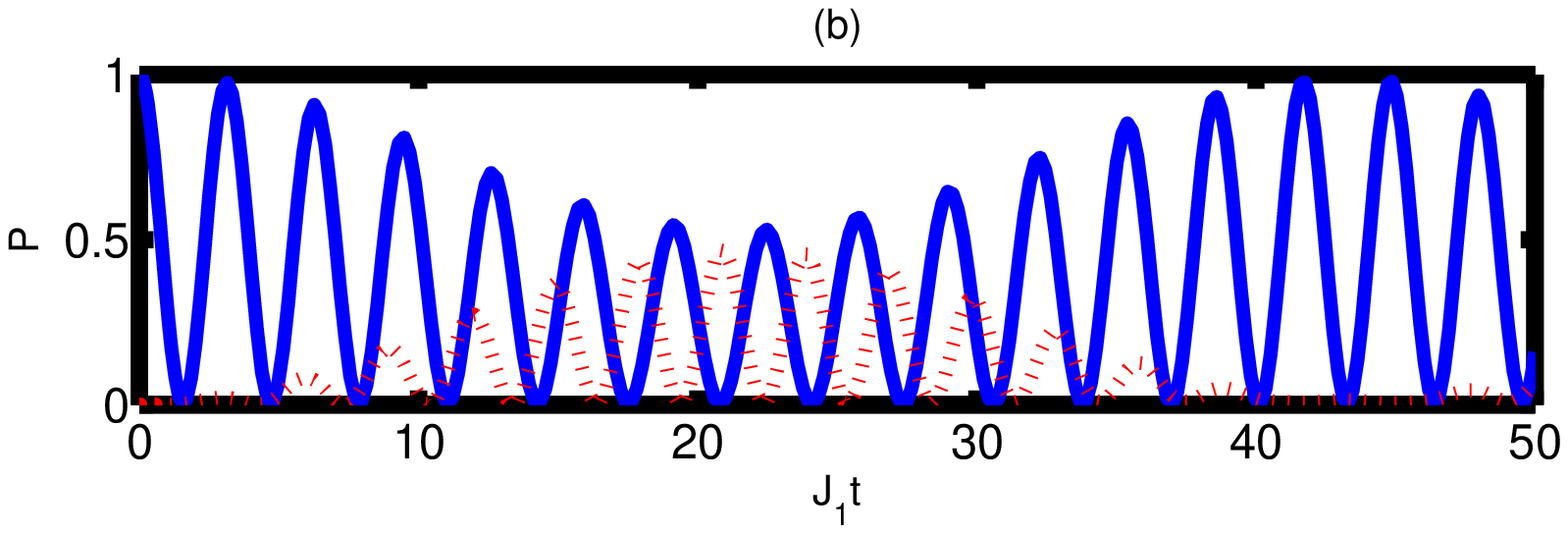}\\
  \includegraphics[scale=0.45]{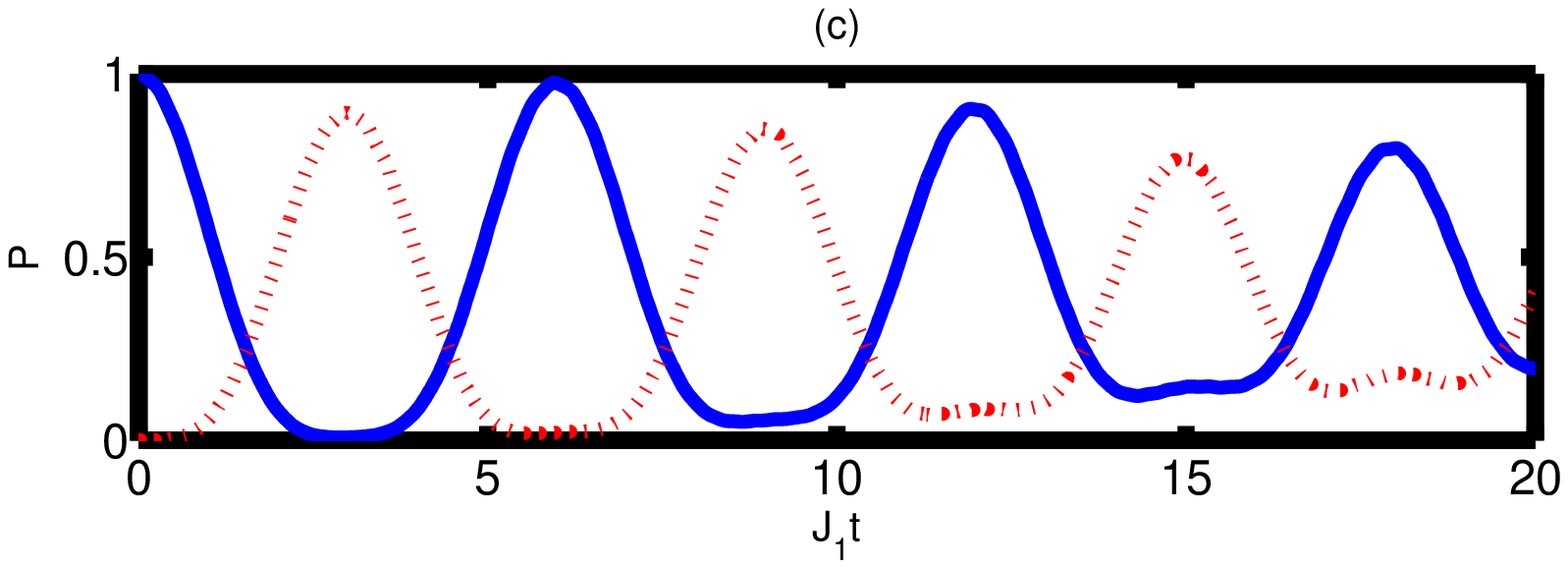}\\
  \caption{(Color online) The time evolution of the populations of the states $|\phi_{1}\rangle$ and $|\phi_{4}\rangle$ for the unbalanced coupling: (a) $\lambda_{1}=0.9J_{1},~\lambda_{2}=0.9J_{2},~J_{2}=0.9J_{1}$; (b) $J_{1}=10\lambda_{1},~J_{2}=10\lambda_{2},~\lambda_{1}=0.9\lambda_{2}$; (c) $J_{1}=0.1\lambda_{1}, J_{2}=0.1\lambda_{2}, \lambda_{1}=0.9\lambda_{2}$. Solid-blue lines represent the population $|D_{1}(t)|^{2}$ of the state $|\phi_{1}\rangle$, while red-dash lines indicate the the population $|D_{4}(t)|^{2}$ of the state $|\phi_{4}\rangle$.}\label{33}
\end{figure}

\begin{figure}
  \centering
  \includegraphics[scale=0.4]{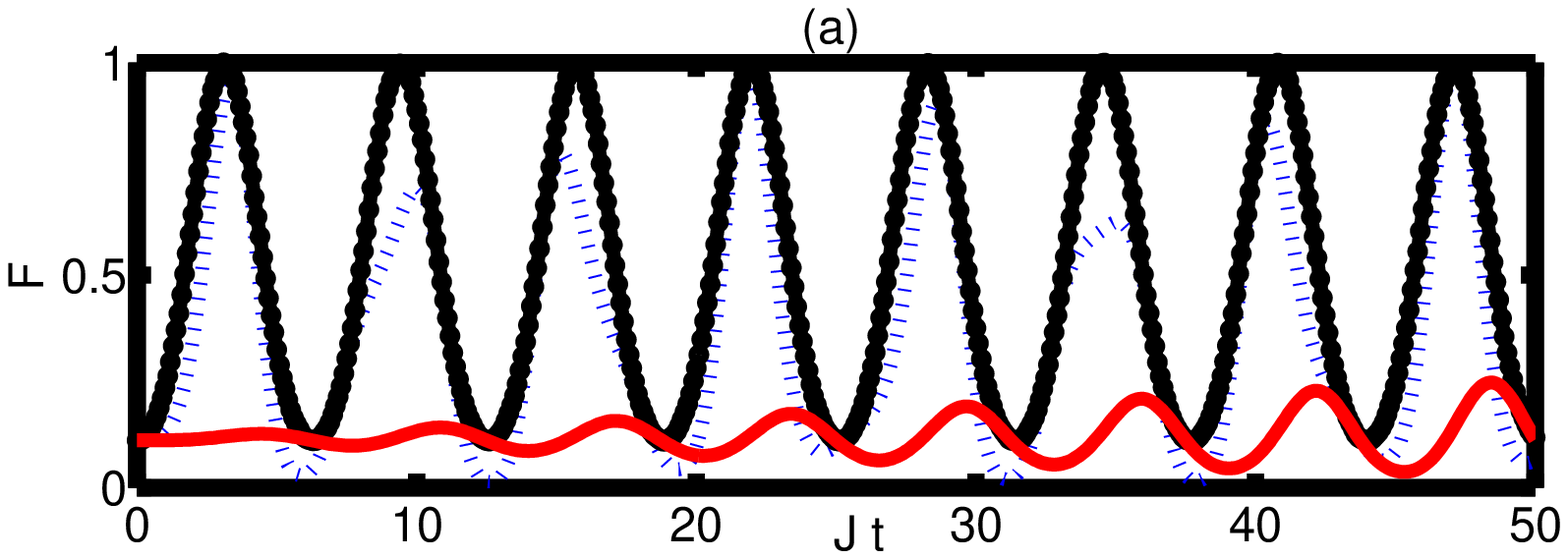}\\
  \includegraphics[scale=0.4]{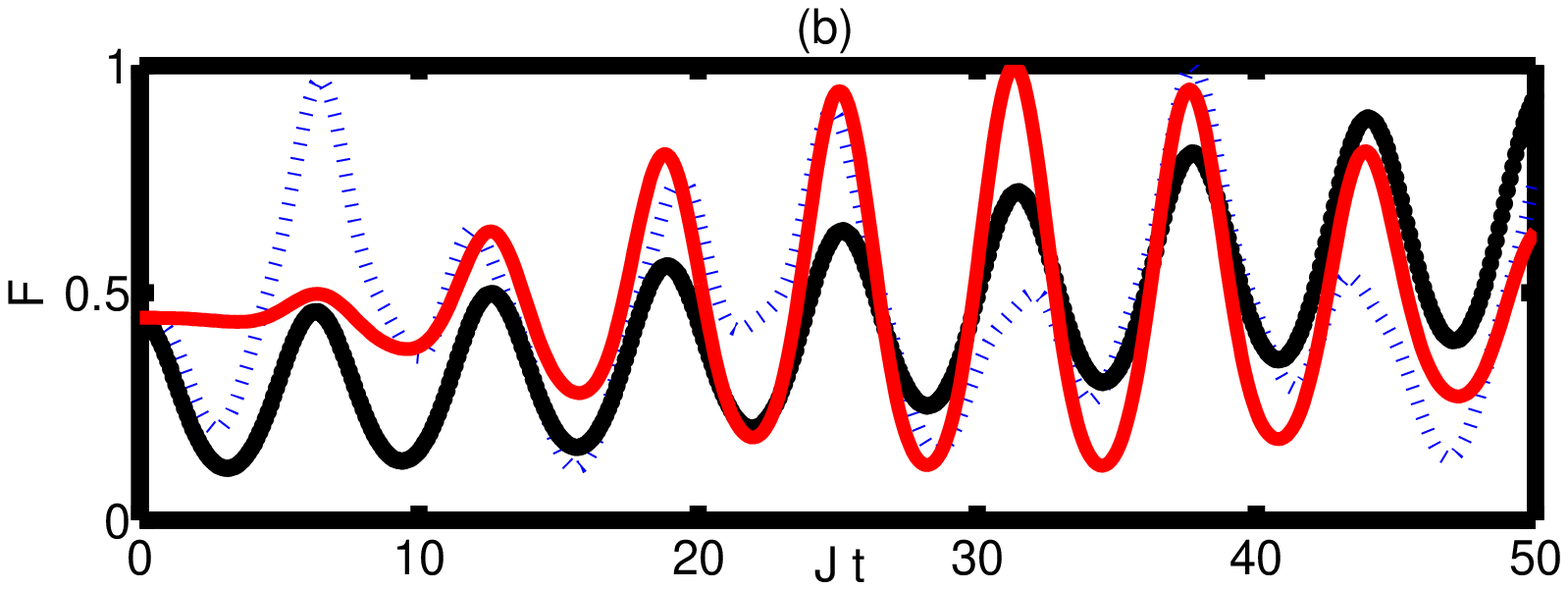}\\
  \caption{(Color online) Fidelity $F$ versus $Jt$, without considering the dissipation of the system. (a) Resonant interaction case. Dashed-blue, solid-dot-black, and solid-red curves correspond to $J=g$, $J=0.1g$, and $J=10g$, respectively. (b) Large detuning case. Dashed-blue, solid-black, and solid-red lines correspond to $J=\lambda$, $J=0.1\lambda$, and $J=10\lambda$, respectively.}\label{4}
\end{figure}

Fidelity is a direct measure to characterize how accurate the QIT is achieved. Here, the fidelity is defined as $F=|\langle\Psi_{T}|\Psi(t)\rangle|^{2}$, where $|\Psi_{T}\rangle$ is the ideal target state of the transfer. The expression of the ideal target state is $|\Psi_{T}\rangle=|g\rangle_{NE_{1}}|0\rangle_{1}|0\rangle_{2}|0\rangle_{L}(\alpha|g\rangle+\beta|e\rangle)_{NE_{2}}$ for the resonant interaction, while $|\Psi_{T}\rangle=|g\rangle_{NE_{1}}|0\rangle_{1}|0\rangle_{2}(\alpha|g\rangle+\beta|e\rangle)_{NE_{2}}$ for the large detuning case. We obtain the expression of the fidelity $F=|\alpha|^{2}+|\beta C_{5}(t)|^{2}$ for the resonant interaction, while $F=|\alpha|^{2}+|\beta D_{4}(t)|^{2}$ for the large detuning case. As an example, let's consider $\alpha=1/\sqrt{3}$ and $\beta=\sqrt{2/3}$. We have $F=(1+2|C_{5}(t)|^{2})/3$ for the resonant interaction case, while $F=(1+2|D_{4}(t)|^{2})/3$ for the large detuning case. The fidelities for the two cases are plotted in Fig. \ref{4}, which shows that high-fidelity QIT between the two NVCEs can be achieved at some moment.

It is meaning to investigating the influence of decoherence of the system on the QIT. When the dissipation of the system is considered, the
dynamics of the lossy system is governed by the following master equation
\begin{eqnarray}
\dot{\rho}&=&-i[H_{I}, \rho]+\frac{\kappa}{2}(2a\rho a^{\dag}-a^{\dag}a\rho-\rho a^{\dag}a)\nonumber\\
&&+\sum_{j}[\frac{\gamma'_{qj}}{2}(\sigma_{z}^{j}\rho\sigma_{z}^{j}-\rho)+\frac{\gamma'_{Nj}}{2}(S_{z}^{j}\rho S_{z}^{j}-\rho)\nonumber\\
&&+\frac{\gamma_{qj}}{2}(2\sigma_{j}^{-}\rho\sigma_{j}^{+}-\rho\sigma_{j}^{+}\sigma_{j}^{-}-\sigma_{j}^{+}\sigma_{j}^{-}\rho)\nonumber\\
&&+\frac{\gamma_{Nj}}{2}(2S_{j}^{-}\rho S_{j}^{+}-\rho S_{j}^{+}S_{j}^{-}-S_{j}^{+}S_{j}^{-}\rho)],
\end{eqnarray}
for the resonant interaction. Here, $\kappa$ is the decay rate of the \emph{LC} circuit, $\gamma'_{qj}$ ($\gamma'_{Nj}$) is the dephasing rate of the \emph{j}-th flux qubit (NVCE), and $\gamma_{qj}$ ($\gamma_{Nj}$) is the relaxation rate of the \emph{j}-th flux qubit (NVCE). For the large detuning case, the \emph{LC} circuit has been adiabatic eliminated in Hamiltonian (\ref{b}). Thus, the master equation is given by
\begin{eqnarray}
\dot{\rho}&=&-i[H_{eff}, \rho]+\sum_{j}[\frac{\gamma'_{qj}}{2}(\sigma_{z}^{j}\rho\sigma_{z}^{j}-\rho)+\frac{\gamma'_{Nj}}{2}(S_{z}^{j}\rho S_{z}^{j}-\rho)\nonumber\\
&&+\frac{\gamma_{qj}}{2}(2\sigma_{j}^{-}\rho\sigma_{j}^{+}-
\rho\sigma_{j}^{+}\sigma_{j}^{-}-\sigma_{j}^{+}\sigma_{j}^{-}\rho)\nonumber\\
&&+\frac{\gamma_{Nj}}{2}(2S_{j}^{-}\rho S_{j}^{+}-\rho S_{j}^{+}S_{j}^{-}-S_{j}^{+}S_{j}^{-}\rho)].
\end{eqnarray}
\begin{figure}
  \centering
  \includegraphics[scale=0.4]{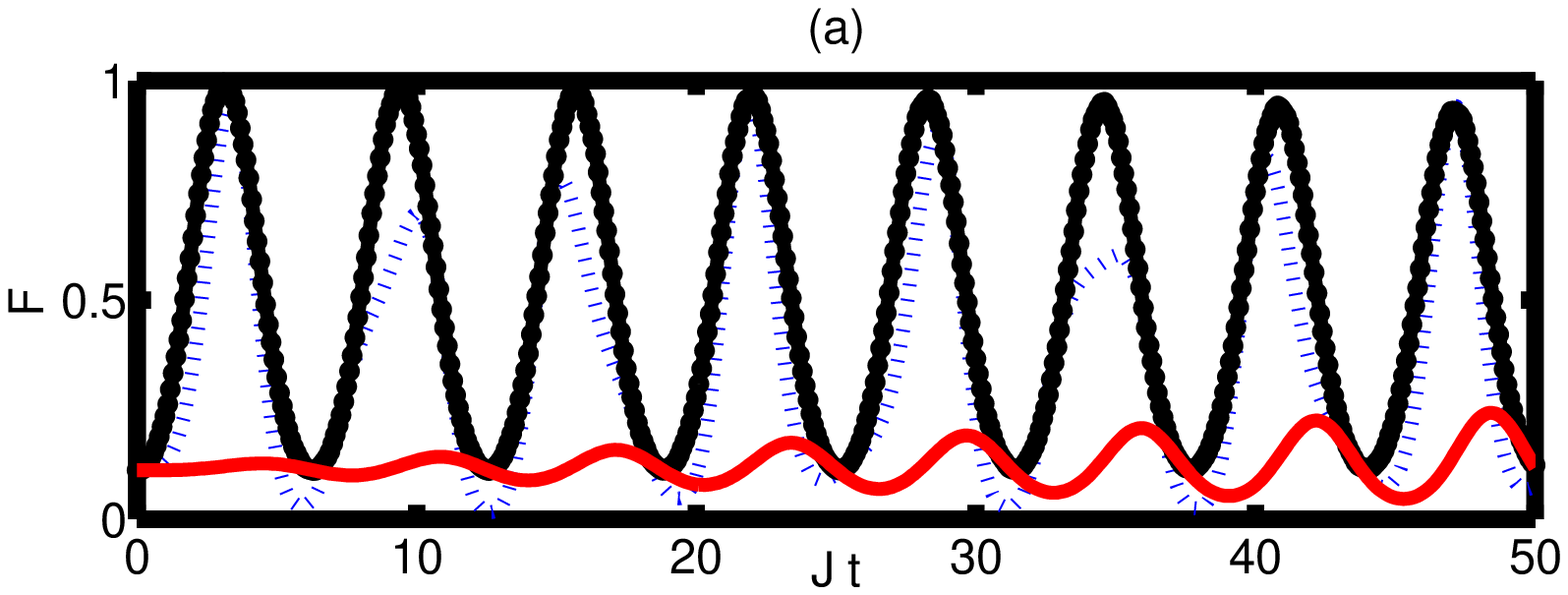}\\
  \includegraphics[scale=0.4]{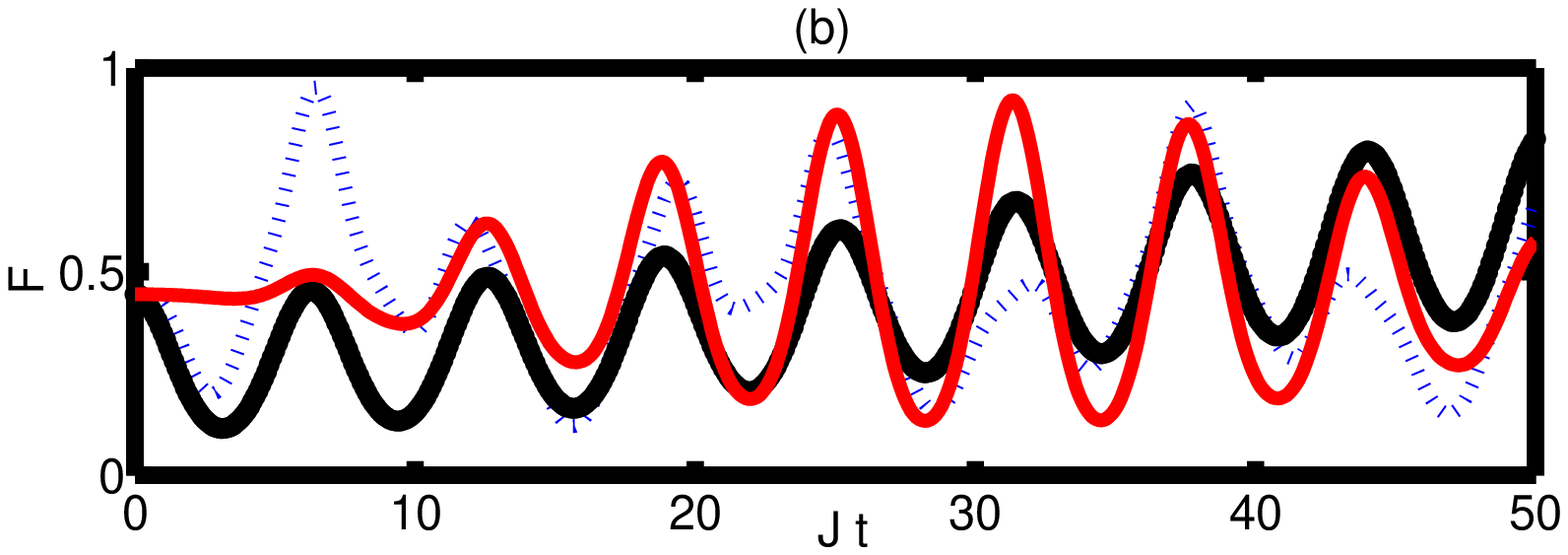}\\
  \caption{(Color online) Fidelity $F$ versus $Jt$, after taking the system dissipation into account. (a) Resonant interaction case. Dashed-blue, solid-dot-black, and solid-red curves correspond to $J=g$, $J=0.1g$, and $J=10g$, respectively. The plot was drawn by setting $\kappa=\gamma'_{qj}=\gamma'_{Nj}=\gamma_{qj}=\gamma_{Nj}=0.001J$. (b) Large detuning case. Dashed-blue, solid-black, and solid-red lines correspond to $J=\lambda$, $J=0.1\lambda$, and $J=10\lambda$, respectively. The plot was drawn by setting $\gamma'_{qj}=\gamma'_{Nj}=\gamma_{qj}=\gamma_{Nj}=0.001J$.}\label{44}
\end{figure}
The Fig. \ref{44} shows fidelity of the QIT versus $Jt$, after taking the dissipation of the system into account. Comparing Fig. \ref{44} with Fig. \ref{4}, one can see that the influence of the dissipation of the system on the fidelity is negligible at small $Jt$.

Let us briefly discuss the experimental feasibility of our proposal. During the annealing process, a NVCE is created by ion implantation into a diamond. A diamond crystal is bonded on top of the flux qubit chip with its surface facing the chip \cite{a6, a7}. The size of a flux qubit is $1 \mu$m order \cite{a5}. A physical system with multiple flux qubits coupled to a \emph{LC} circuit has been proposed \cite{a11}. The time $t$ of the QIT is inverse ratio to the coupling strength $J$. For the coupling strength $J\approx70$MHz, we have $t\sim 1/J\sim14$ ns, which is much shorter than the flux qubit's coherence time $T_{2}\simeq20 \mu$s \cite{h} and the NVCE's coherence time approaching $1$ second \cite{h1}. Also, decoherence of the flux qubits and the NVCEs can be effectively suppressed by periodic dynamical decoupling \cite{j}.

\section{Extending to the scalable quantum circuit}

\begin{figure}
  \centering
  \includegraphics[scale=0.45]{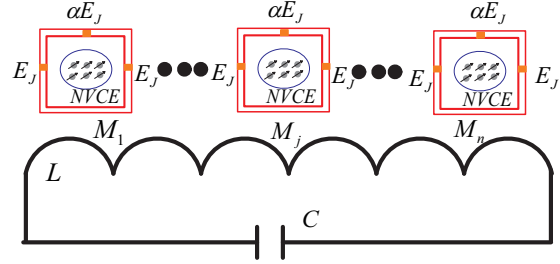}\\
  \caption{(Color online) Scalable quantum information transfer circuit. Multiple flux qubits are coupled to a \emph{LC} circuit by their mutual inductances $M_{j}(j=1,2,...n)$. Each flux qubit couples to a NVCE acting as an information memory unit.}\label{5}
\end{figure}
Our proposal can be extended to the scalable quantum circuit, which is constructed by $n$ flux qubits, NVCEs and a \emph{LC} circuit acting as a data bus, shown in Fig. \ref{5}. All flux qubits can be made to be coupled (or decoupled) with the \emph{LC} circuit by varying the external flux applied to each qubit loop. Alternatively, one can replace the small junction of each flux qubit with a SQUID and change external magnetic field threading the SQUID loop \cite{f0, f, k}, such that each flux qubit is coupled or decoupled to the \emph{LC} circuit. In this way, the information can be transferred between any two selected NVCEs. Furthermore, the architecture provides the possibility for creating entanglement among NVCEs and performing quantum logic operations on NVCEs, which are important in quantum information processing.

\section{Conclusion}
A hybrid architecture has been proposed for realizing QIT between NVCEs. For both resonant interaction and large detuning cases, it has been explicitly shown that high-fidelity QIT can be achieved between two spatially-separated NVCEs, and is robust against decoherence of the hybrid architecture. Also, a discussion has been given for the influence of the different coupling mechanisms on the QIT. According to the current experimental conditions, the feasibility of this proposed has been analyzed. The proposed architecture opens a way for scalable QIT among NVCEs, which is important in large scale quantum information processing. Finally, the method presented here is applicable to a wide range of
physical implementation with different types of data buses such as nanomechanical
resonators and TLRs.

\begin{acknowledgments}
FYZ thanks Prof. Chong Li and Dr. Bao Liu for valuable discussions. FYZ and HSS were supported by the
National Science Foundation of China under Grants No. 11175033. ZFY was supported by the National Science Foundation of China under Grants Nos. 11447135 and 11447134, and the Fundamental Research Funds for the Central Universities No. DC201502080407. CPY was supported in part
by the National Natural Science Foundation of China under Grant Nos. 11074062 and 11374083, the Zhejiang Natural Science Foundation under Grant No. LZ13A040002, and the funds from Hangzhou Normal University under Grant Nos. HSQK0081 and PD13002004. This work was also supported by the funds from Hangzhou City for the Hangzhou-City Quantum information and Quantum Optics Innovation Research Team.
\end{acknowledgments}

\end{CJK*}
\end{document}